\documentstyle[tighten,aps,prb,epsfig]{revtex}

\begin{document}
\draft
\preprint{LA-UR-99-3361}

\title{\bf Physics of rapidly expanding supercritical solutions: A
first approach}
\author{ Shirish M. Chitanvis}

\address{Theoretical Division, MS B268\\
Los Alamos National Laboratory, Los Alamos, New Mexico 87545}
\date{\today}
\maketitle
\begin{abstract}
We consider the case when a supercritical fluid emerges at sonic speed 
from a small orifice in a high pressure chamber.
The subsequent expansion causes a pressure drop and the fluid then
enters a regime where its equation of state in $P-V$ space becomes
concave towards the origin.
This is the signal for an expansion shock to occur in a non-ideal fluid.
This paper provides the details of an analytic calculation of the
shape
and location of this
expansion shock using Whitham's front-tracking method.
Dependence of the shape of the front on various operating conditions
was calculated for the particular case of supercritical carbon dioxide.
The results shed
light on the rapid expansion of supercritical solutions (RESS),
a process which is used in many manufacturing technologies. 

\end{abstract}


\newpage
\section{Introduction}

The use of supercritical solutions to enhance processes in the
chemical industry dates back several decades.
Their utility stems from the fact that supercritical fluids have a
relatively high diffusivity, which promotes mixing on the microscopic
level, and their density is high, allowing relatively high
solubilities\cite{mchugh}. 
Furthermore, the ability to selectively dissolve substances by varying 
thermodynamic parameters like pressure and temperature is also
a big reason why supercritical solutions are used in chemical
processing. In addition, the environmentally benign nature of some
supercritical solvents, e.g. carbon dioxide, encourages their use in industry.

In many applications, such as pharmaceutical processing or 
ceramics processing, rapid expansion of supercritical solutions (RESS) 
has been promoted due to the control available on the morphology
of the end-product.
For example, delivery of drugs within the human body may be enhanced
if the chemicals coalesce to a certain particle size\cite{mchugh}.
In ceramics processing, fine control over particle size may lead to
products with enhanced strength\cite{matsen}.  

Surprisingly, despite its industrial importance\cite{mchugh}, little
effort appears to have been devoted to 
understanding the dynamics of the RESS process.
This process typically consists of expanding a supercritical
solution, which is under pressure, through a capillary-sized nozzle
into the ambient atmosphere.
Depending on the nozzle design, the fluid velocity as it exits the
nozzle into the ambient atmosphere can be either sub-sonic or supersonic.
In fact, unless the nozzle is designed carefully, the exit velocity
will be sub-sonic\cite{ss}, in which case the subsequent flow will be
subsonic and begin slowing down in the expanding geometry.
This can lead to a complicated flow, where the downstream state can
influence the flow upstream into the nozzle region.
We therefore intend to focus on those processes where the 
fluid exits the nozzle at sonic speed.
In this case, the fluid accelerates as it exits the nozzle, and 
its pressure drops
until vaporization sets in.  This
causes the solute to begin precipitating out in the form of
micron-sized particles. 
The flow then decelerates as the liquid transforms into a multi-phase
spray and drag effects set in.
Currently, empiricism is used to determine maximal
operating conditions for the RESS process.
While this approach may be a practical way to solve the problems of
the day, we point out that the RESS process, despite its
apparent simplicity, involves an enormously large range of physics.
This physics has largely remained unexplored.
In fact, the characteristic information
(or lack thereof) conveyed by photographs of the
process\cite{matsen}is a {\it fuzzy} conical plume of 
presumably fine particles emanating from the orifice of a high
pressure chamber, and this provides us minimal knowledge of the
basic phenomena which lead to the precipitation of the solute.
There are indeed experimental papers that study evaporation waves  
in liquids, but the focus of those investigations has been
super-heated, {\it sub}-critical fluids\cite{simoes,sturtevant}, rather 
than supercritical fluids, which we consider here.
It is hoped that our paper will spawn experimental efforts in the
supercritical fluid
arena, from which verification of our theoretical results may be
sought.

For conceptual ease, we can divide the physical problem into stages:

\begin{description}
\item[ $\bullet$  ]~ First of all the solvent, which is in the
  supercritical state, 
undergoes a transition to a vapor-like state when the fluid enters an
expansion regime in which its equation of state in $P-V$ space ($P$ is the
pressure, $V=1/\rho$, $\rho$ being the density) becomes 
concave towards the origin\cite{bethe}.
This transition is a rarefaction or expansion shock.
As the solvent vaporizes, the solute begins to precipitate.
Thus, it is important to know the location and shape of this  
expansion shock, as it is the driving mechanism for the subsequent
precipitation of solid particles.

\item[ $\bullet$  ]~ Secondly, the RESS process
should be viewed more as an {\it anti-detonation wave}, that we refer to
here as a {\it vaporization wave}.
In this vaporization wave, heat of formation is taken away from 
the mechanical motion as the liquid transforms into vapor.
While the process of the solvent transforming to a vapor is a relatively 
fast process, the subsequent mechanism of particle formation may take
place over a spatially extended zone.
The extent of the vaporization wave depends strongly on the processes
that can transport heat
in and out of this thin zone from both the fluid and the gas, as well
as the details of the nucleation and aggregation process.
For a stable operation, the expansion
shock in our case must be stationary.

Furthermore: 

\item[ $\bullet$  ]~ Just as detonation waves have modes of
instability\cite{erpenbeck}, it is 
important to investigate the stability of this vaporization
wave.
This aspect of the study is quite important, as it will affect the
subsequent production of particles.

\item[ $\bullet$  ]~ Depending on the local flow conditions, the
expansion shock front produced in the expanding flow
can lead to the formation of a Mach disk.
This Mach configuration will be different from the usual ones in that
we have at hand colliding oblique expansion shocks, rather than the
usual compressive strong shocks\cite{shirish2}.

\end{description}

Since the initial conditions that define the morphology of the
precipitated solid are set in the region where the dynamic phase
transition occurs from a condensed to a vapor-like phase, it is
important to be able to model the complex 
physics of supercritical fluid expansion well. 
As far as we are aware, there is no 
computational fluid dynamics (CFD) code which contains 
a comprehensive physics package capable of providing insight
into the phenomena we wish to investigate. This combined with
problems of resolving the thinness of the multi-phase vaporization
zone suggests that another approach is desirable.

To begin our investigation of this wide range of phenomena,
we shall focus in this paper on the calculation of the shape and
location of the expansion shock in the case of a pure solvent
(supercritical carbon dioxide), leaving for future work the theory of
how the solute precipitates beyond this expansion shock.
And we shall examine briefly the stability of the expansion shock.

Expansion shocks are a relatively
poorly studied phenomenon compared to compressive shocks.
Their thermodynamics 
were first studied by Hans Bethe about fifty years
ago\cite{bethe,menikoff,cramer1,kluwick}.
The behavior of the so-called Bethe-Zeldovich-Thomson
(BZT)\cite{thomson} fluids has been studied by a few researchers in the
past, and most of the work has, until recently, focused on
one-dimensional problems\cite{cramer1,kluwick}.
Only a few investigations have been reported in two
dimensions\cite{eu,cramer2,argrow,japan}. 
And of these, none have reported on the particular geometry which is
most likely to be used in the RESS process, viz., a pinhole in a high
pressure chamber.
Our goal is to present an analytic approach to this problem, which we
believe sheds useful light on the RESS process.
It also lays the groundwork for future numerical work which must
follow, and which will hopefully eliminate some of the approximations
which we have employed.

In order to track the shape and location of the expansion shock, we
propose to use level-set methods,  which have 
proven their efficacy in problems involving the propagation of
detonation waves in chemical explosives\cite{bdzil1,Chen,Osher}.
It is interesting to note that a precursor of this front
tracking technique was considered several decades ago by Hans
Bethe\cite{bethe2}, during his investigations of the atomic blast wave.


\section{The physical criterion for an expansion shock in a flowing
supercritical fluid} 

Let us begin by considering the specific geometry of the {\it nozzle}
through which the supercritical fluid is emerging.
In an ideal situation, one would consider a deLaval nozzle, with a
smoothly varying converging portion, connected at the neck to a
smoothly expanding channel.
Such nozzles have been considered in the past, with the main result
that for the ideal case when the fluid goes sonic at the throat of the 
nozzle, there is subsequent expansion, and an expansion shock occurs
some distance past the throat, in the expanding portion of the nozzle\cite{eu}.

In most practical cases, it is more likely that the nozzle is
basically a pinhole (orifice) in a high pressure chamber filled with the
supercritical fluid, so that the interior walls of the chamber leading up to
the orifice can be considered as a converging nozzle, while the
exterior walls represent the {\it expanding}
portion of the nozzle.
The fluid exits through the orifice.
Our plan is to compute the flow of supercritical
carbon dioxide just {\it outside} this orifice.
It is always possible in principle to achieve a state where the Mach number 
is precisely unity exiting the orifice.
When this occurs, the subsequent flow outside the orifice (to be thought of
as a flow in a diverging channel) is supersonic.
It is this particular case of supersonic flow past the orifice which
we will consider here.
In the case of sub-sonic flow exiting the orifice, the problem is
actually more difficult, since the upstream and downstream states of
the fluid can communicate via sound waves, whereas, for supersonic
flow, for obvious reasons, this difficulty is not present.

As the supercritical fluid exits the orifice at Mach one, it speeds up, expands,
and as the pressure drops, it enters a regime where
the equation of state becomes concave.
As Bethe showed many years ago,
this is the signal that an expansion shock can occur.
Across the shock, the fluid goes from a condensed supercritical state
to a vapor-like state.
Such a case is denoted schematically in Fig.1, where two
adiabats for supercritical 
carbon dioxide are displayed.
The expansion shock is denoted schematically as a transition from (0) 
to (1).
The physical model we have chosen is that the fluid undergoes an
expansion shock as soon as it enters the concave portion of the
equation of state (0).
We shall assume that the shock is sufficiently strong that state (1)
lies in {\it regular} convex region.
The state (1) could lie in the concave region if the shock was
sufficiently weak.
It would then 
decay further via subsequent expansion shock(s) into lower density
states.

To calculate the adiabats, we used Van der
Waal's equation of state (VdW EOS).
The VdW EOS is a cubic EOS, and
represents the simplest way of describing a first order phase
transition. 
We can write down analytically
its adiabatic form:

\begin{equation}
\left(P + {a\over V^2}\right)~\left(V-b\right)^{R/C_v+1} = constant
\label{VdW}
\end{equation}

where $V = 1/\rho$, $\rho$ is the density, $P$ is the pressure,
$a$ and $b$ are the usual Van der Waal's parameters, $R = 3.814\times
10^7~(ergs-mole^{-1}-^{\circ}K^{-1})$ is
the universal gas constant, 
and $C_v$ is the specific heat, which is quite large near the critical 
point ($\sim 50 R$).
For carbon dioxide, in c.g.s. units,
$a = 3.959\times 10^{12} \mu^{-2}$,
$b = 42.69 \mu^{-1}$,
where $\mu=44$ is the molecular weight of carbon dioxide.
The critical pressure for carbon dioxide is $73.8~bars$, and the critical
temperature is $31 ^{\circ}C$.

The actual flashing/vaporization event will be
described by Chapman-Jouguet jump conditions.
For our problem, the Chapman-Jouguet jump conditions describe an {\it
  anti-detonation}, in which energy must be supplied to the molecules
to break bonds and form vapor.


\section{Prandtl-Meyer flow near the orifice}

We shall begin our calculation of the shape and location of the
vaporization front by considering how the supercritical fluid, flowing 
at supersonic velocities, expands just as it emerges from the orifice.
The next two figures show schematically the geometry we used.
The first of these simply shows the orifice, and the second {\it zooms} in 
on the half space we considered in the immediate vicinity of the
corner representing the exit at the orifice.

Thus this looks just like the Prandtl-Meyer flow problem, but applied
to a non-ideal fluid (supercritical carbon dioxide) obeying a cubic EOS.
As is done for the Prandtl-Meyer problem, we shall use cylindrical
geometry, with the axis of the cylinder coming out of the paper, and
the only independent variable is the azimuthal angle $\phi$ which
describes the turning of the flow around the corner.
The standard equations which need to be solved to obtain 
the profile of the pressure versus the turning angle are given below:

\begin{eqnarray}
{d u_r(\phi) \over d \phi} &&= a(\rho(\phi)) \nonumber\\
\rho(\phi) a(\rho(\phi)) \left({d a(\rho(\phi)) \over d \phi} +
  u_r(\phi) \right) &&= -{d P(\rho(\phi)) \over d \phi}
\label{pmarr}
\end{eqnarray}

where $u_r$ is the radial component of the velocity, $a$ is the
adiabatic speed
of sound, $P$ is the pressure, given by the adiabatic form of the
equation of state stated earlier.
As is done conventionally, we have assumed that near the corner, the
hydrodynamic variables depend only on the azimuthal angle $\phi$.

The pressure profile is given in Fig. 4.
The dashed line in Fig. 4 indicates the maximum turning angle possible,
corresponding to the entrance of the fluid into the concave portion of
the equation of state, as discussed above.

We notice two interesting facts about the state of supercritical
carbon dioxide as it turns the corner:

{$\bullet$} For the case of a polytropic gas, the Prandtl-Meyer fan
terminates 
at the angle for which the pressure goes to zero.
However, we are considering a condensed fluid, which can undergo an
expansion shock as it enters the concave portion of the equation 
of state, before it can achieve negative pressure.

{$\bullet$} Secondly, the maximum turning angle is close to $90$
degrees. 
This turning angle is consistent with a
nearly spherical flash front. 

As we increase the pressure at which supercritical carbon dioxide leaves 
the orifice, the maximum angle of turning before an expansion shock
can occur begins to decrease.
This maximum turning angle will also decrease if we consider that the
fluid might have a radial component of the velocity, so that the
speed of the fluid exiting the orifice has Mach number greater than
one, and it expands at a faster rate.
Such a possibility could be achieved by flaring the exit of the
orifice.
 
The next step will be to take this Prandtl-Meyer type of calculation,
which applies only in  
the immediate neighborhood of the corner, and extend it using 
Whitham's front-tracking technique to distances farther from the
corner we just looked at, in order to get a more complete picture of
the flash front.


\section {An introduction to Whitham's front-tracking method.}

We have a supercritical fluid emerging from an orifice
into a much larger space at Mach one.
As this fluid speeds up, it expands, and flashes to a vapor state
across some surface which we shall refer to as the vaporization front.
In this paper we shall study the initial transition from a
supercritical fluid which has just entered the concave region of the
equation of state, to a vapor-like state which lies in a convex region.
There are other cases (weaker shocks), which we shall not consider
here, when the 
state across the shock lies in a concave region.
Then there may be an extended zone beyond
in which additional processes take place, such as a decay via
further expansion shocks into a dispersed wave.
This topic will be studied in future investigations.

In the previous section, we described how an extension of the usual
Prandtl-Meyer expansion calculation can be used to describe the
vaporization in the vicinity of the edge of the orifice.
Here, we shall describe how to extend this calculation much beyond the 
edge.
The method we shall employ is the front-tracking method of Whitham\cite{whitham}.
It begins with the intuitive notion that the front is described by a
direction (the normal to the surface), and the {\it density} of rays
associated with this surface.
Whitham derives a conservation law for the {\it ray density}, which
could be thought of as being analogous to a current density, so that:

\begin{equation}
\vec \nabla \cdot \left({\hat n \over A}\right) = 0
\label{cons}
\end{equation}

where $\hat n$ is the unit normal to the surface, and $A$ is the area
(of the stream-tube) associated with it.
One can find an explicit expression for $A$ in the case of spherical
geometry, which is the one we shall use here.
If $\Psi(\vec r,t)=constant$ describes the surface, $\vec r$ being the
independent spatial co-ordinates and $t$ the time, then:

\begin{equation}
\hat n = {\vec \nabla \Psi \over \vert \vec \nabla \Psi \vert}
\label{normal}
\end{equation}

So far, the description is quite general, and we would like to
determine a connection with the physics of vaporization.
In order to do that, it turns out to be convenient to {\it hop} on to the 
liquid emanating from the orifice.
Then, the front, which is stationary in the laboratory frame, will
appear to be moving {\it inwards}, towards the observer sitting on the liquid.
It is then easy to show that the velocity of the front is given by:

\begin{equation}
v_n = -{\partial_t \Psi \over {\vert \vec \nabla \Psi \vert}}
\label{vn}
\end{equation}

where the subscript $n$ denotes the normal to the surface, and $v$ the 
velocity.

Furthermore, we shall follow Whitham and make an asymptotic expansion
of $\Psi$ in powers  
of $t$, and retain only the linear term.
This is done since our main interest is in the $t=0$ limit, when the
location of the front will be taken to coincide with that in the
laboratory frame, as an initial condition.
Additionally, it can be seen that $\Psi$ can be defined arbitrarily to 
within an overall normalizing constant, so we shall write down the
equation to the surface as:

\begin{equation}
\Psi(\vec r,t) \equiv \psi(\vec r) - a t = constant
\label{asymp}
\end{equation}

where $a$ is the speed of sound.
Equation \ref{cons} can now be expressed as:

\begin{equation}
\vec \nabla \cdot \left({M \over A}~\vec \nabla \psi(\vec r)\right) 
= 0
\label{am1}
\end{equation}

where $M$ is the Mach number.
If the physics of vaporization could be
included in some fashion through a relation between $A$ and
$M$, then Eqn.\ref{am1} would become relevant to the vaporization
phenomenon. 
We shall do this through a slight modification of the usual method
which is used for tracking the motion of a compressive shock wave through a
complicated geometry.
It must be noted that the conventional method of making the $A-M$
connection is an
approximation, as is our modification of it\cite{whitham}.
The approximation we make is that of spherically symmetric dynamics.
Recall that we aim to describe the vaporization phenomenon as an
abrupt change in the hydrodynamic variables from the supercritical to the
vapor phase, while preserving mass, momentum and conservation.--
This is of course analogous to an {\it anti-detonation} as described
earlier.
In fact, a version of Whitham's method has been developed to describe
successfully the propagation of detonation fronts in complicated
geometries by Bdzil and co-workers\cite{bdzil1}.
And it is an analogy to this problem we have in mind while trying to
obtain the shape of the vaporization front.

The basic idea is to ask how the hydrodynamic variables behave if the
front is positioned at different locations in the spatial region of
interest.
In order to do this, it is convenient to assume that the geometry is
changing smoothly.
This is appropriate for the problem at hand, and 
leads to a set of duct equations.
These duct equations incorporate the jump conditions for the
vaporization phenomenon, so that they describe approximately how the
Mach number changes as the area $A$ changes.
For completeness, note that for the spherical geometry which we assume
in the vicinity
of the orifice, $ dA(r)/dr = 2 A(r)/r$, r
being the radial variable.

The jump conditions for the vaporization process, when the front is
stationary in the laboratory frame (to ensure a stable operation) are:

\begin{eqnarray}
\rho_v u_v &&= \rho_L u_L \nonumber\\
P_v + \rho_v u_v^2 &&= P_L + \rho_L u_L^2 \nonumber\\
{1\over 2} u_v^2 + h_v &&= {1\over 2} u_L^2 + h_L \nonumber\\
h_j &&= P_j/\rho_j + e_j + g_j,~\forall~j=v,L
\label{hug-cond}
\end{eqnarray}

where the subscript $v$ denotes the vapor phase and $L$ the liquid
phase, $\rho$ is the density, $u$ is the velocity, $e_j$ is the
internal energy and $g_j$ is 
the heat of formation in each of the phases.

We can assume a {\it stiffened-gas} equation of state for this
portion of the calculations viz., $P_L \approx \Gamma_L^{-1}
a^2 \rho_L$, 
where $\Gamma_L \sim 30$
was adjusted to yield the correct pressure in the appropriate density
range.
This equation of state is a
representation of the relation between pressure and density in the
supercritical state in $P-\rho$
space {\it just} before the fluid enters the {\it concave} region
where an expansion shock takes it to a vapor state.
We can make  the approximation that the front is a {\it
strong} expansion shock (vapor density turns out to be a few per cent
of the density in the supercritical state, and that the internal
energy in the vapor 
phase is much lower than that in the liquid phase.
Now, the
jump conditions across the front can be written in a more
convenient form:

\begin{eqnarray}
\rho_v &&\approx  
{{\rho_L M^2} \over {\left( \Gamma_L^{-1} + M^2 \right)}} \nonumber\\ 
u_v &&\approx  
\left({a \over M}\right)~ \left( \Gamma_L^{-1} + M^2 \right)  
\nonumber\\  
P_v &&\approx \alpha \rho_v(M) - {1 \over 2} \rho_L a M u_v(M) 
\nonumber\\  
\alpha &&\approx  
\left( C_{vL} T_L - \Delta + P_L/\rho_L - {1 \over 2} u_v(M)^2 \right) 
\label{CJ}
\end{eqnarray}

where $M$ is the Mach number in the liquid phase, as is the speed of
sound $a$, and $\Delta \equiv g_v - g_L$ is 
the heat of vaporization which must be supplied to the system for
vaporization to occur. 

Upon inserting appropriate numerical values for the various variables, 
it turns out that:

\begin{equation}
\alpha \approx  C_{vL} T_L
\label{approx}
\end{equation}

This is due to the high value of the specific heat near the critical
point, which is the regime of interest in this paper.
In this sense, the $A-M$ relation we shall derive shortly will be
fairly insensitive to most of the physical parameters in the problem.

As mentioned earlier, a successful application of Whitham's technique
requires a moving 
front, whereas for reasons of ensuring a stable operation, we took
the front to be stationary in Eqn.(\ref{CJ}).
If we go to the frame of reference in which the fluid emanating from
the orifice is stationary, then the front will possess a velocity $-u_L$.
In this frame of reference, the phase front which separates the
supercritical 
fluid from the vapor state will appear to be moving inwards, towards
the observer.
In fact, it will also appear to the observer that the orifice itself
is shrinking.
Note that in going to the frame of reference that is
moving with the supercritical fluid, $u_v \to u_v - u_L \approx u_v$,
so that the jump conditions we used earlier remain approximately the same.
As indicated earlier, our goal is to concentrate on the short-time
limit, so that the consequences of being in a reference frame such
that the orifice is shrinking to an infinitesimal value can be avoided.

Using Whitham's ideas for the propagation of a front along a {\it
  duct} of slowly varying cross-section, we can get a differential
equation relating 
the change in $M$ to the change in $A$.
To do this, we can use his characteristic equation which describes the 
propagation of a jump discontinuity along a duct.
It is easy to see that his derivation holds for our problem (an
anti-detonation), which 
differs from the one discussed in his book for the propagation of a
regular shock wave.
It is important to note that the derivation is valid in the limit of
short times, when changes the geometry as the front propagates are
small.
In this linearized regime:

\begin{equation}
{dP_v\over dr} + \rho_L a {du_v \over dr} + \left({\rho_L a^2 u_L \over
  u_L + a} \right) {dlnA \over dr} = 0
\label{char}
\end{equation}

Inserting the appropriate equations from Eqn.\ref{CJ}, we obtain:

\begin{equation}
{dM \over dr} \left[ {2 \alpha M \over a^2 \Gamma_L
\left(\Gamma_L^{-1} + M^2 \right)^2} - M +1 -{1 \over \Gamma_L^{-1}
M^2 }\right]~ 
\left({M+1 \over M}\right) = {-1 \over A(r)} {dA(r) \over dr}
\label{am2}
\end{equation}

It is implicit in this equation that the front does not deviate
excessively from a spherical shape.
This comes from the fact that we are using duct equations, which
assume small deviations from the basic symmetry of the problem, in
this case, spherical symmetry.
For the problem at hand, ours is the first calculation to explore the
shape and location of the vaporization front, and as such will provide a mark 
for future, more sophisticated calculations to compare against.
More importantly, it is hoped that these calculations will spur
experimental work in this area, perhaps using schlieren techniques,
etc. which will undoubtedly provide useful information regarding the
phenomenon.  

Using $ \Gamma_L^{-1} << 1$, it is possible to deduce, using a Taylor
expansion, that for $r \sim r_o$, $r_o$ being the orifice radius: 

\begin{eqnarray}
M(r) &&\approx b^{-1} \left({r_o^2 \over r^2} -1 \right) +1 \nonumber\\
{A(r_o) M(r) \over A(r)} &&\approx b^{-1} + (1-b^{-1}) {r_o^2 \over
  r^2} \equiv f_{\Delta}(r) \nonumber\\
b &&=  {4 \alpha(\Delta) \over {\Gamma_L a^2}}
\label{mr}
\end{eqnarray}

It turns out that for the parameters relevant to the problem at hand,
$b^{-1} << 1$, because its value 
dominated by the internal energy term. 
This in turn implies an insensitivity to the other parameters
defining the problem, as discussed earlier.

Note that it might appear that the physics of the expansion shock is now
represented by the single 
parameter $b$, and in particular by its (negligible) dependence on
$\Delta$, the heat of vaporization, through the parameter $\alpha$.
Actually, this is too simplistic a view.
The expansion shock, and the geometry, are also directly
represented by the 
imposition of the condition that initially, the
front is {\it hinged} to the corner of the orifice at an angle close
to $90^{\circ}$ (see the previous section).

The reason we have restricted attention to $r \sim r_o$ is as follows.
One may guess that since the front is anchored
to the edge of the orifice, and almost parallel 
to the axis (i.e. has an approximately spherical shape in that
vicinity), that the rest of the front will be approximately spherical
as well.
It will be shown later that this assumption is well-founded for a wide
parameter range.

The problem is now reduced to solving:

\begin{equation}
\vec \nabla \cdot \left( f_{\Delta}(r) \vec \nabla \psi(\vec r) \right)=0
\label{fe1}
\end{equation}

In this sense Eqn.\ref{mr}, which is really an $A-M$ relation
expressed in terms of the radial variable, represents a simplification 
of the problem, in that we now have to {\it merely} solve a linear
partial differential equation, as opposed to a fully non-linear partial
differential equation. 
We shall next attempt a solution of this equation by means of a
separation of variables  
in spherical co-ordinates viz., $r,\theta$, assuming that the orifice
is azimuthally symmetric, and the origin being at the center of the
orifice, with the z-axis coinciding with the axis of symmetry.


\begin{section}{Fundamental solutions of the front-tracking equation.}

We will now obtain the linearly independent solutions to the following 
partial differential equation, derived in the previous section.
We shall set the origin at the center of the orifice, and use
spherical co-ordinates, with the z-axis coinciding with the symmetry
axis.

\begin{equation}
\vec \nabla \cdot \left( f_{\Delta}(r) \vec \nabla \psi(\vec r) \right)=0
\label{fe1}
\end{equation}

A separation of variables ($r,\theta$) turns out to be successful.
In other words, setting $\psi(\vec r) = P_{\nu}(cos(\theta)) R_{\nu}(r)$, where
$P_{\nu}$ is a Legendre function ($\nu$ does not have to be an
integer), leads to the following equation for $R_{\nu}(r)$:  

\begin{equation}
R_{\nu}''(r) + \left[ {2 \over r} + {f'_{\Delta}(r) \over f_{\Delta}(r)}
\right] R_{\nu}'(r)  + {\nu (\nu + 1) \over r^2} R_{\nu}(r) = 0
\label{rnu}
\end{equation}

where a prime indicates a derivative with respect to the appropriate
independent variable.
Note that we do not require $\nu$ to be an integer, as we shall use
$\nu$ as a parameter chosen to fit a certain boundary condition.
This is similar to solving electrostatics problems in
sharp corners, etc.
In this case, $P_{\nu}(cos(\theta))$ is no longer a polynomial, but
an infinite series.

The radial equation can be solved using {\it Mathematica}.
The answer is in terms
of the Hypergeometric function, the two linearly independent solutions 
being:

\begin{eqnarray}
R_{\nu}^{(1)}(r) &&=
r^{(3-\sqrt{9-\epsilon(\nu)})/2}~_2F_1({\cal A}_1, {\cal
  B}_1, {\cal C}_1,-{b \beta r^2 \over r_o^2}) \nonumber\\  
{\cal A}_1 &&={1\over 2}-{1 \over 4} \sqrt{1-\epsilon(\nu)}-{1 \over
4} \sqrt{9-\epsilon(\nu)} \nonumber\\ 
{\cal B}_1 &&= {1\over 2}+{1 \over 4} \sqrt{1-\epsilon(\nu)}-{1 \over
4} \sqrt{9-\epsilon(\nu)} \nonumber\\ 
{\cal C}_1 &&= 1-{1\over 2} \sqrt{9-\epsilon(\nu)} \nonumber\\
R_{\nu}^{(2)}(r) &&=
r^{(3+\sqrt{9-\epsilon(\nu)})/2}~_2F_1({\cal A}_2, {\cal
  B}_2; {\cal C}_2;-{b \beta r^2 \over r_o^2}) \nonumber\\  
{\cal A}_2 &&= {1\over 2}-{1\over 4} \sqrt{1-\epsilon(\nu)}+{1\over 4}
\sqrt{9-\epsilon(\nu)} \nonumber\\ 
{\cal B}_2 &&= {1\over 2}+{1\over 4} \sqrt{1-\epsilon(\nu)}+{1\over 4}
\sqrt{9-\epsilon(\nu)} \nonumber\\ 
{\cal C}_2 &&=1+{1\over 2} \sqrt{9-\epsilon(\nu)}\nonumber\\
\epsilon(\nu) &&= {4 \nu (\nu+1)} \nonumber\\
\beta &&= \left(1-b^{-1}\right)
\label{fund}
\end{eqnarray}

One requires regularity of the solution at the origin on physical
grounds, in order to obtain as stable a solution as possible. 
The first solution diverges at the origin when $\epsilon(\nu) < 0$.
The sign depends on the value of $\nu$.
Now the value of $\nu$ is determined by a boundary condition
(discussed below).
If we now demand that a {\it strong} condition of regularity
for {\it all} values of $\nu$ be satisfied, then one must reject the
first solution. 
Furthermore, when we used the first solution $R_{\nu}^{(1)}$ by itself, we
could not find a value of $\nu$ which would satisfy the boundary condition
discussed below, for a vaporization angle close to $90^{\circ}$.

The second solution $R_{\nu}^{(2)}$ is regular at the origin.
It would be physically 
acceptable, as long we can satisfy the condition that at the initial
moment, the normal to the front is hinged at
$\theta\equiv\phi_f$ to the axis of symmetry.
We do this by demanding that for $r=r_o,\theta=\pi/2$:

\begin{equation}
 {r_o R_{\nu}'(r_o) P_{\nu}(\pi/2) \over {
     R_{\nu}(r_o)
      \left[dP_{\nu}(Cos(\theta))/d\theta\right]_{\theta=\pi/2}}}=
  -tan(\phi_f)  
\label{hinge}
\end{equation}

It is through this equation (\ref{hinge}) that the presence of the
corner at the junction of the orifice and the tunnel is acknowledged.
We will show in the next section that the shape of the front depends
on the value of $\phi_f$. 

The condition given by Eqn.\ref{hinge} is to be thought of as an
eigenvalue problem. 
The question of normalization of the solution remains.
But this is trivial, since we can choose the normalization to be arbitrary, 
as a change in the normalization constant leaves the differential
equation for the front and Eqn.\ref{hinge} unchanged.
This arbitrary choice is equivalent to using an arbitrary choice of
the temporal origin.


\section {Calculating the shape and location of the vaporization 
front}  

The maximum value of the vaporization angle $\phi_f$ turns out to be close
to $1.54~radians$, as discussed earlier, for the case that the fluid
emerges from an orifice in a high pressure chamber at a pressure of
$79 bars$ (just above the critical pressure), at Mach one.

It turns out that to satisfy Eqn.(\ref{hinge}), $\nu
\approx {\cal O}(10^{-2})$ when $\phi_f \sim 1.54$.
To locate the front, we compute the {\it contour} corresponding
to the  
value of $\psi(\vec r)$ at $\vec r \equiv (r_o,\pi/2)$, knowing that
at that point in space, Eqn.(\ref{hinge}) is satisfied.
Remember that the normalization of the solution is arbitrary, and as
pointed out earlier, this corresponds to choosing an arbitrary
temporal origin.

The next figure (Fig.5) shows the locus of roots of the equation
$\psi(\vec r) = \psi(r_o,\pi/2)$, for $\phi_f = 1.54$.

We will now consider what happens as the exit velocity has a different 
value than the one just considered.
The primary reason for doing this is that it may eventually impact the 
morphology of the precipitate.
We shall do this by assuming that during the computation of the
Prandtl-Meyer fan in the previous section, the exit velocity possesses 
a radial component.
This could be achieved in principle by flaring the orifice smoothly a short
distance before the sharp corner is reached.
Then
the fluid will undergo an expansion shock at smaller values of the
angle $\phi_f$. 
We shall now simply study the shape of the vaporization front
parametrically as $\phi_f$ is decreased.
This serves the additional purpose of allowing us to study the
sensitivity of the flash front to $\phi_f$, and thus has a bearing on
the stability of the front to such perturbations.

The next two figures (Figs. 6 and 7) show what happens to the shape and location of 
the front as we decrease $\phi_f$.
At $\phi_f \approx 1.0$, the front is clearly seen to start
flattening near the axis.

The flattening becomes more pronounced as we decrease $\phi_f$
further.
Note that as this occurs, the front begins to deviate increasingly
from its hemispherical shape.
We thus begin to violate the assumption of spherical symmetry
$f_{\Delta}(\vec r) \equiv f_{\Delta}(r)$ more and 
more as $\phi_f$ decreases.
For smaller values of $\phi_f$, the calculated front shows dramatic
departures from sphericity.
Future
investigations may wish to focus on these cases to probe the implications
of the change of the shape of the front for the underlying fluid flow
and the subsequent precipitation process.
At the moment, our calculation is the most sophisticated one available 
for the geometry considered, and the extreme case we now consider
(Fig. 7) may
be compared against future numerical calculations of the
front for this geometry.

\end{section}

\section{Stability of the expansion shock}

We adapted Whitham's front-tracking method to our steady state case by 
changing the reference frame appropriately.
In this frame, the expansion shock is seen to converge inwards,
towards the orifice.
We then have a converging front sweeping across a fluid,
converting it from a condensed state to a vapor state as it moves.
In this sense our problem is analogous to a converging detonation
wave, the main difference being that the heat of vaporization in our
case is negligible, whereas heat release is the main feature of
detonations. 
It is thus natural to ask whether the expansion shock we have
calculated is stable to perturbations.

Let us begin by recalling the arguments given by Whitham\cite{whitham} 
regarding the stability of converging shock waves in spherical
geometry.
Whitham shows that his front-tracking technique provides a solution
which diverges at the origin. 
Thus, any perturbation (described in terms of spherical harmonics)
gets amplified as the wave converges towards the center.
In our case, the analog is that $P_{\nu}(cos(\theta))$, the angular
component of the solution diverges at $\theta=\pi/2$.
It is therefore tempting to claim that the expansion shock is
unstable. 
However, our interest is not in the convergence of the shock towards
the orifice, but {\it solely} in the $t=0$ limit, so that we can
obtain the shape and location of the (stationary) shock in the
laboratory frame.
Therefore, from the discussion in the previous section, we see that
the solution remains finite for the physical range of interest in
$(r,\theta)$ space, even if we perturb the boundary condition by
varying the angle $\phi_f$ at which the fluid undergoes an expansion
shock at the edge of the orifice.
In this sense the expansion shock we have computed is stable.

\section{Conclusions}

Progress has been made towards understanding the RESS process by first 
making a conceptual connection with the phenomenon of expansion shocks 
predicted many years ago by Hans Bethe.
Whitham's front-tracking method was then adapted to calculate the
location and shape of the expansion shock in a supercritical fluid
emerging from a pinhole in a high pressure chamber at Mach one (so
that the subsequent flow is supersonic).
For this case the front is fairly hemispherical in shape.
We presented other cases, when the fluid emerges from the orifice
directly at supersonic speeds, where the deviation from sphericity
becomes increasingly dramatic.
While these extreme cases violate the underlying assumption of
spherical symmetry made initially, we speculate that they may be
correct in a qualitative sense.
This point must be verified by future analyses which eliminate the
assumptions made in this paper.
It is also hoped that our paper will spawn experimental efforts in this
arena, from which verification of our theoretical results may be
sought.


\section{Acknowledgments}

I would like to acknowledge useful discussions with John Bdzil, Tariq
Aslam, and especially Robert Owczarek regarding the front-tracking
formalism.
I would like to thank Larry Hill and Ralph Menikoff for their comments.
This research was
supported by the Department of Energy, under contract W-7405-ENG-36.




\begin{figure}
\caption{
 Two adiabats for supercritical carbon dioxide are shown 
  in this figure.  The expansion shock is denoted schematically by
  the arrow from (0) to (1).  The state (0) denotes the point on the
  adiabat where the EOS changes from being convex to concave.
}
\label{fig1}
\end{figure}
\begin{figure}
\caption{
  A schematic of the geometry, showing a high pressure
  chamber with an orifice in it.  The direction of flow and the
  symmetry axis are indicated.}
\label{fig2}
\end{figure}
\begin{figure}
\caption{
  This figure zooms in on the half-space utilized in the 
  Prandtl-Meyer calculation of the supercritical carbon dioxide going
  around a sharp corner.
}
\label{fig3}
\end{figure}
\begin{figure}
\caption{
  The pressure profile for supercritical carbon dioxide
  going around a sharp corner, as obtained from a Prandtl-Meyer
  calculation for a fluid obeying a Van der Waal's equation of state.
}
\label{fig4}
\end{figure}
\begin{figure}
\caption{
  The vaporization front is almost hemispherical for the
  case $\phi_f = 1.54$.
}
\label{fig5}
\end{figure}
\begin{figure}
\caption{
  The vaporization front is less hemispherical for the
  case $\phi_f = 1.0$.
}
\label{fig6}
\end{figure}
\begin{figure}
\caption{
  This figure denotes a dramatic departure from
  sphericity when $\phi_f = 0.3$.  This case violates the underlying
  assumption of spherical symmetry in the calculation. We speculate
  that the result may be valid in a qualitative sense.}
\label{fig7}
\end{figure}


\newpage
\centerline{ S.M. Chitanvis,\ \ Fig.\ \ref{fig1}}

\begin{figure}[t]
\epsfig{file=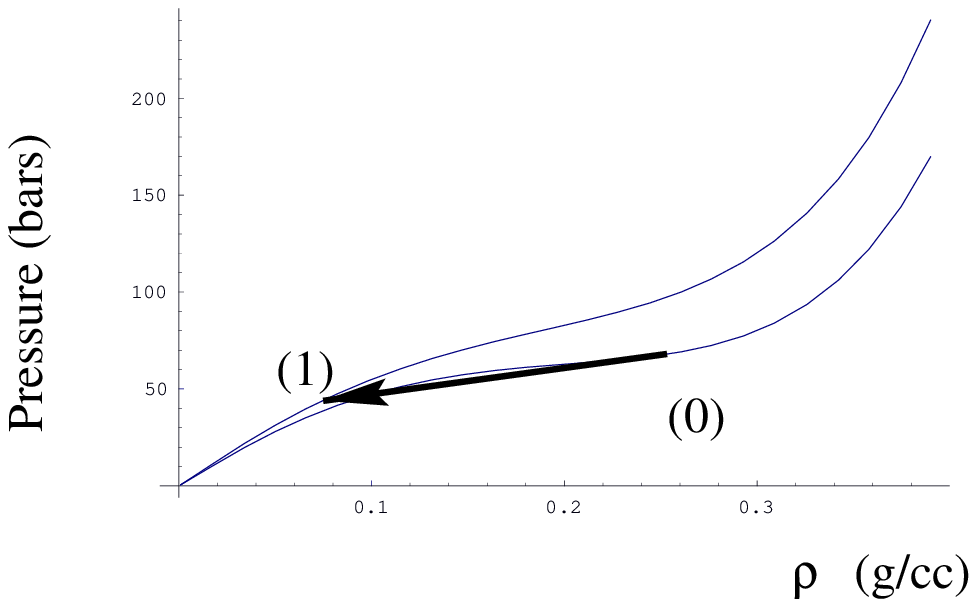,width=8cm}
\end{figure}


\centerline{ S.M. Chitanvis,\ \ Fig.\ \ref{fig2}}
\begin{figure}[t]
\epsfig{file=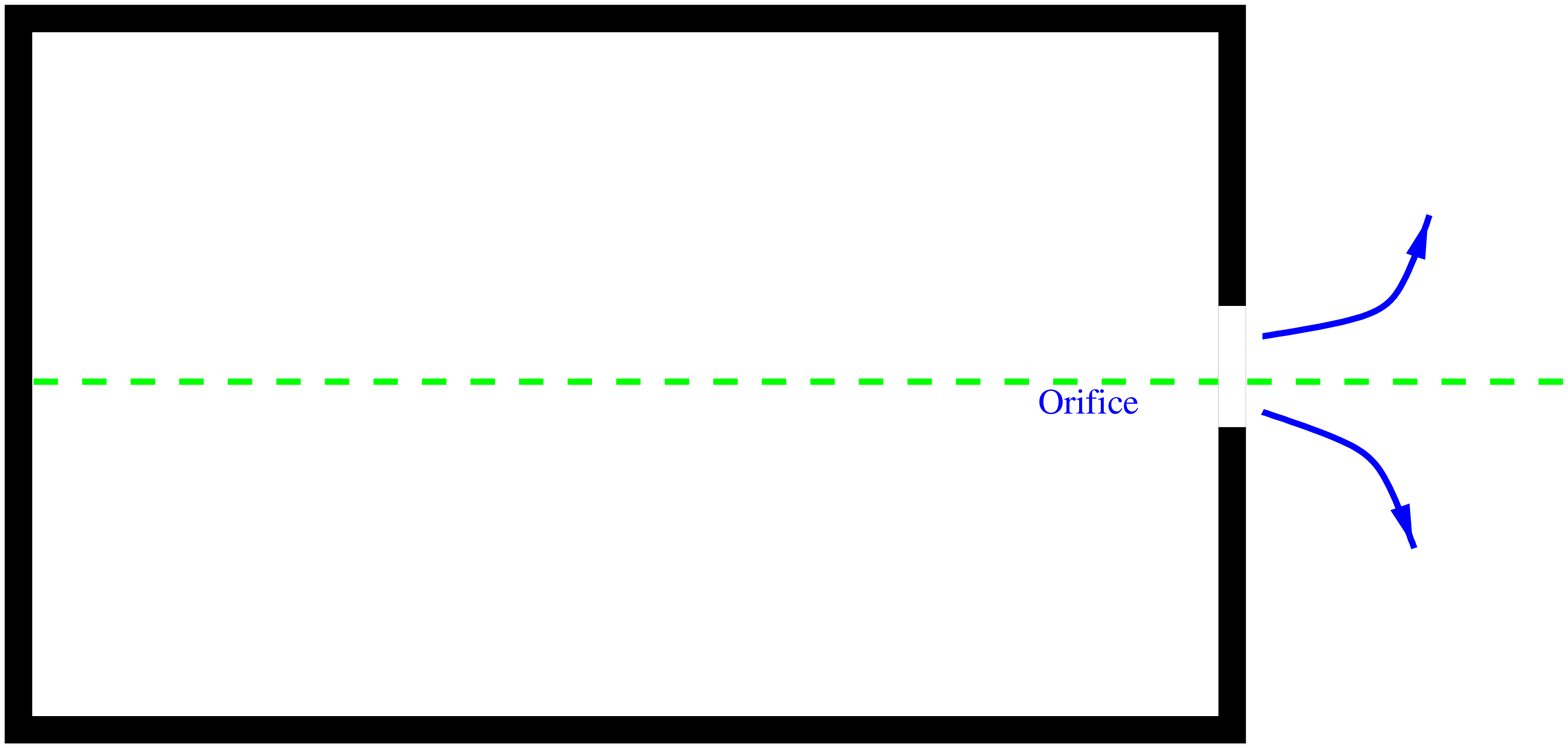,width=8cm}
\end{figure}


\centerline{ S.M. Chitanvis,\ \ Fig.\ \ref{fig3}}
\begin{figure}[t]
\epsfig{file=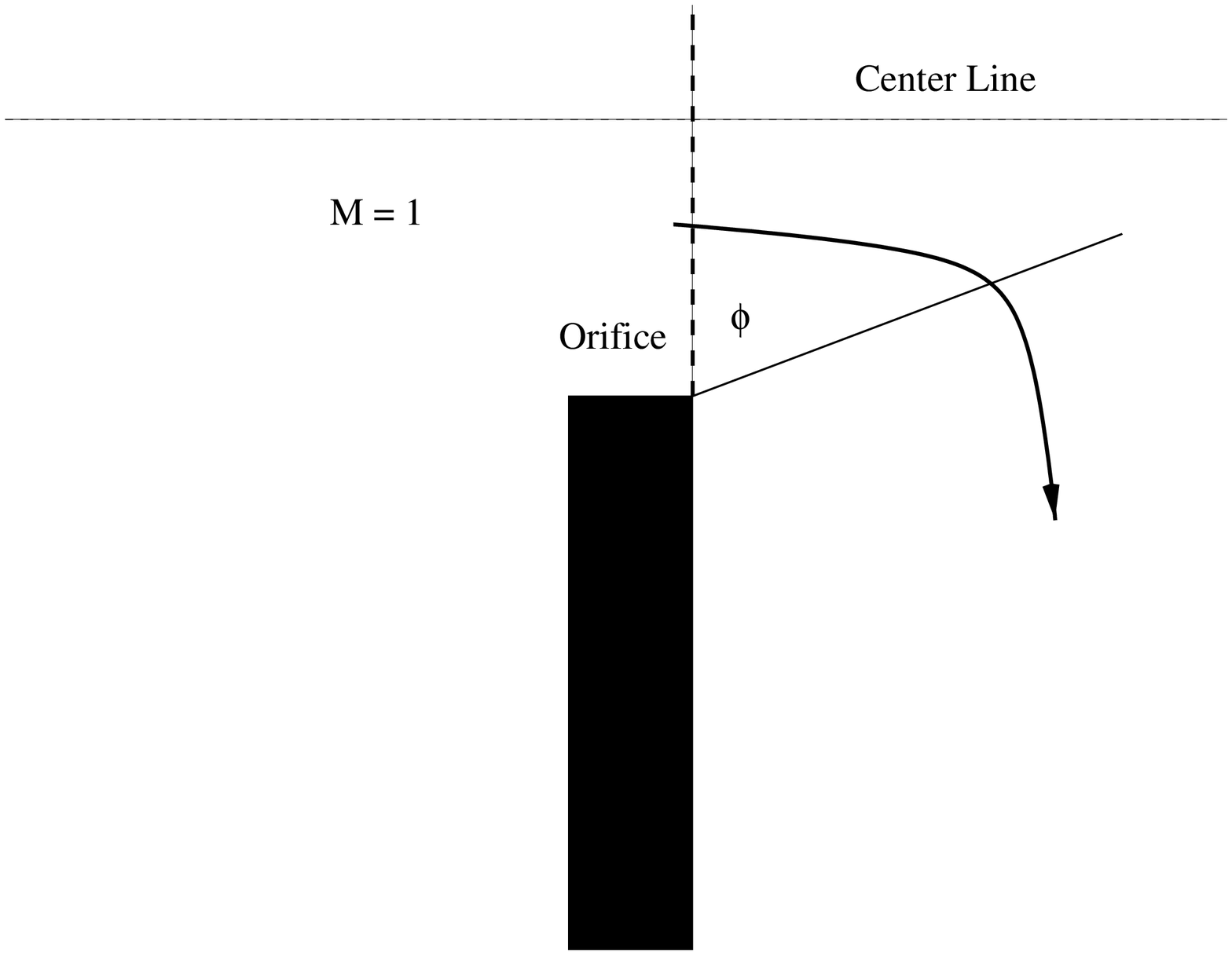,width=8cm}
\end{figure}


\centerline{ S.M. Chitanvis,\ \ Fig.\ \ref{fig4}}
\begin{figure}[t]
\epsfig{file=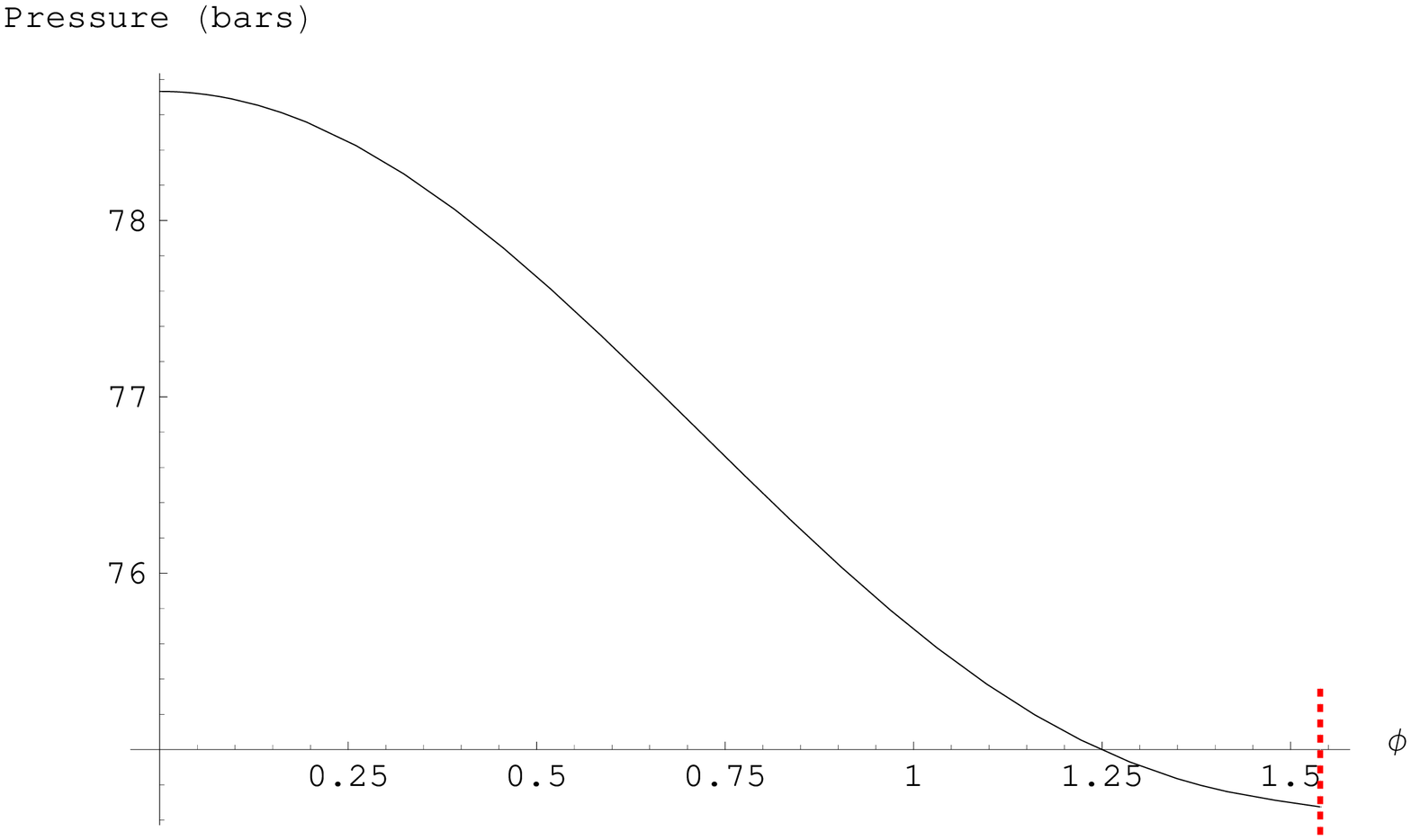,width=8cm}
\end{figure}


\centerline{ S.M. Chitanvis,\ \ Fig.\ \ref{fig5}}
\begin{figure}[t]
\epsfig{file=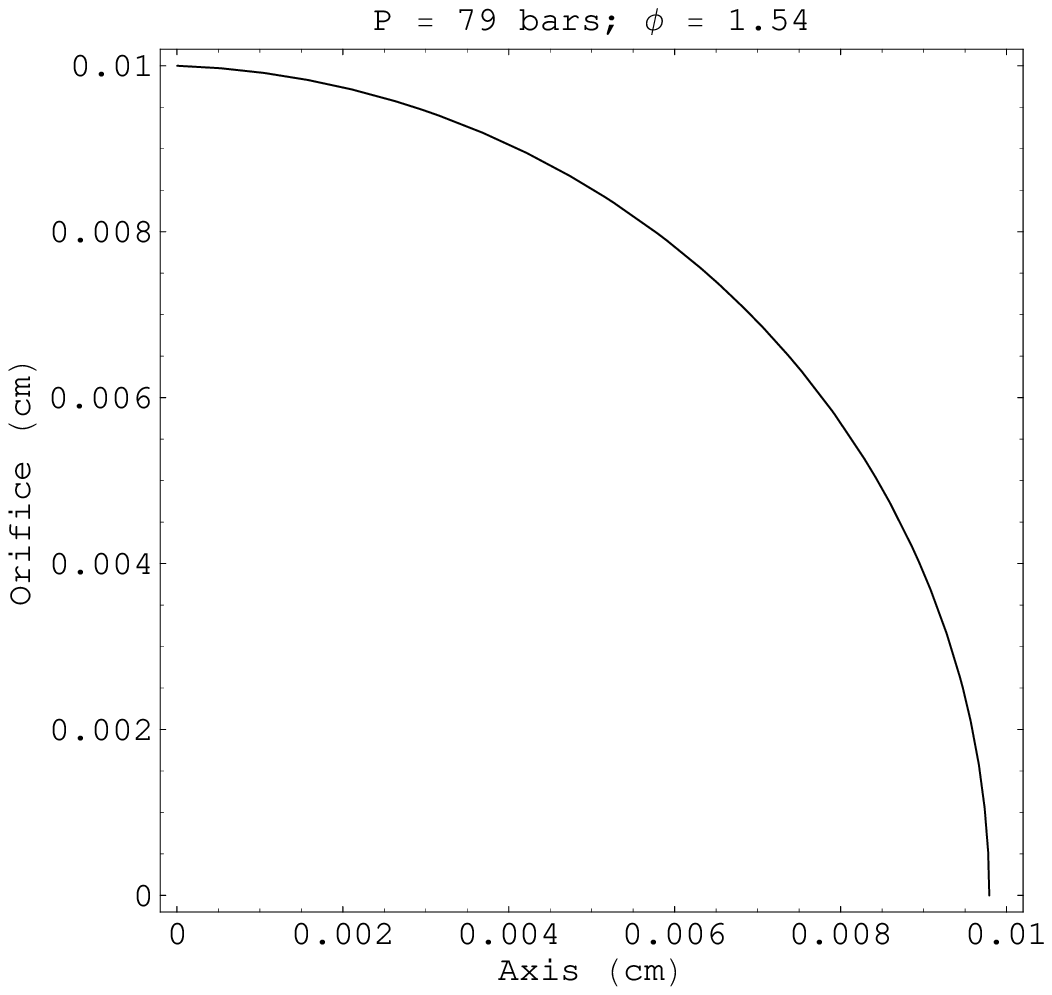,width=8cm}
\end{figure}


\centerline{ S.M. Chitanvis,\ \ Fig.\ \ref{fig6}}
\begin{figure}[t]
\epsfig{file=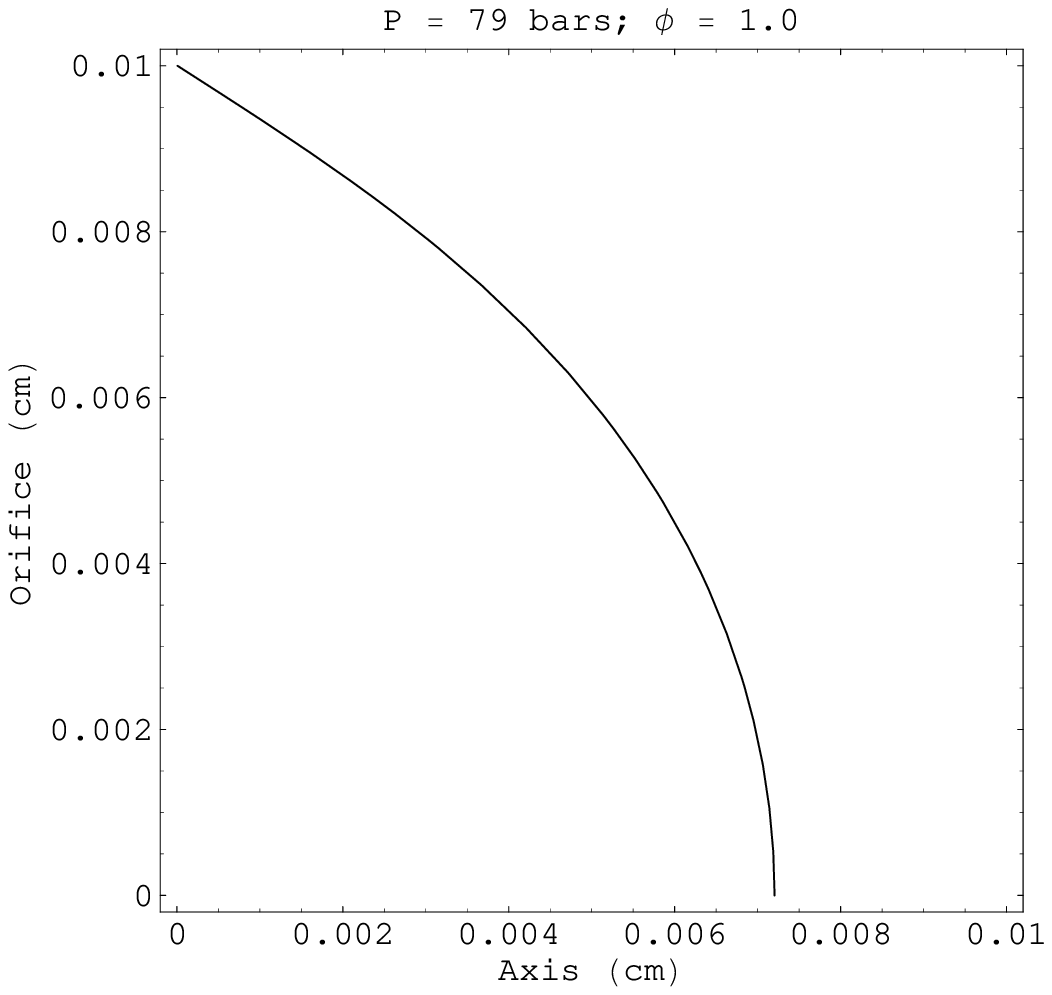,width=8cm}
\end{figure}

\centerline{ S.M. Chitanvis,\ \ Fig.\ \ref{fig7}}
\begin{figure}[t]
\epsfig{file=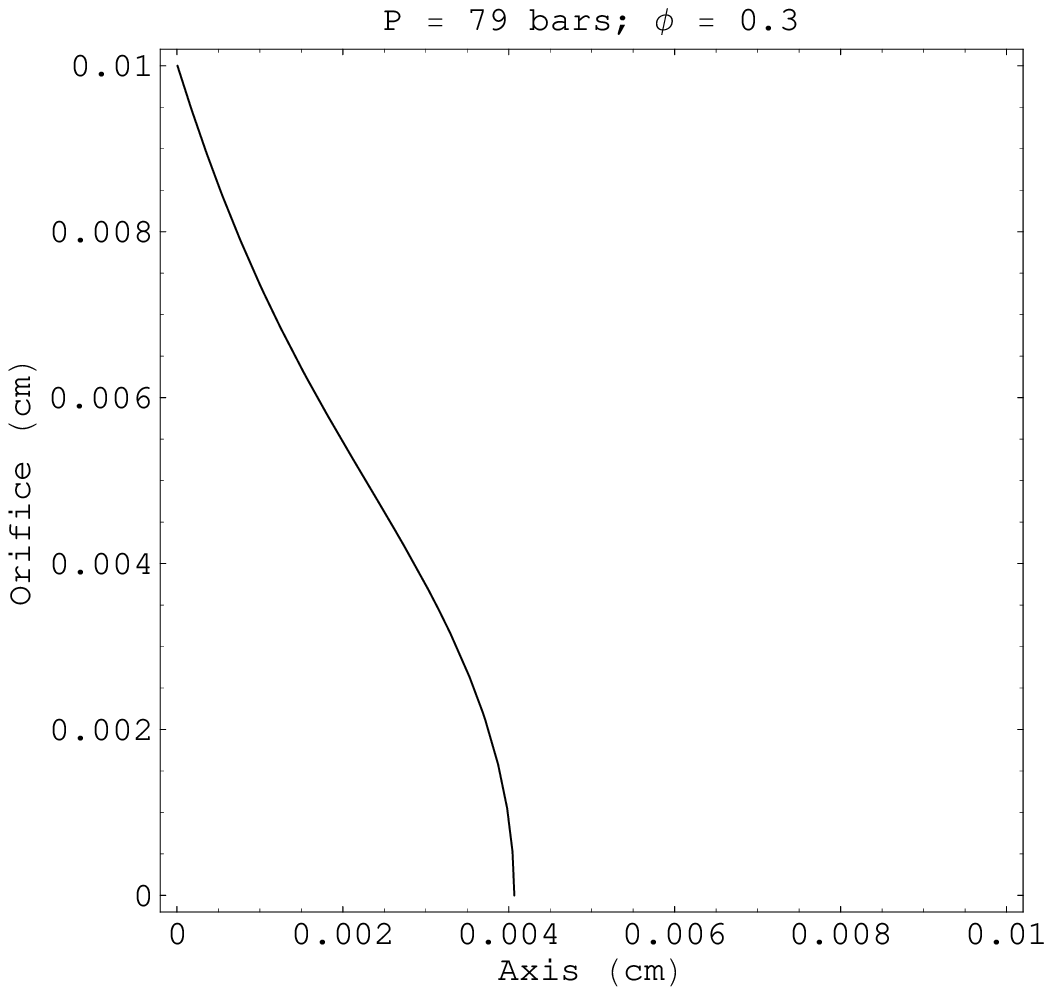,width=8cm}
\end{figure}

\end{document}